\documentclass{elsart}

\usepackage{amsmath}
\usepackage{amssymb,amsfonts}
\usepackage{epsfig}
\usepackage{psfrag}
\journal{Acta Materialia}

\newcommand{\vc}[1]{{\boldsymbol #1}}
\newcommand{\bw}[1]{\raisebox{1.5ex}[-1.5ex]{#1}} 
\newcommand{\mub}{\mu_{\mbox{\tiny{B}}}}
\newcommand{\mev}{\small{\frac{\mbox{meV}}{\mbox{atom}}}}
\newcommand{\sub}[1]{\mbox{\tiny{#1}}}

\begin{document}

\begin{frontmatter}



\title{First-principles investigation of the Ni-Fe-Al system}


\author[Stuttgart]{F. Lechermann\thanksref{new}}
\ead{Frank.Lechermann@cpht.polytechnique.fr}, 
\author[Stuttgart]{M. F\"{a}hnle} and
\author[Austin]{J.M. Sanchez}
\address[Stuttgart]{Max-Planck-Institut f\"{u}r Metallforschung, 
Heisenbergstrasse 3, D-70569 Stuttgart, Germany}
\address[Austin]{Texas Materials Institute, The University of Texas at Austin,
Austin, Texas, 78712, USA}
\thanks[new]{present address: CPHT \'{E}cole Polytechnique, 91128 Palaiseau 
Cedex, France.}
\begin{abstract}
By combining ab-initio electron theory and statistical mechanics, the 
physical properties of the ternary intermetallic system Ni-Fe-Al in the 
ground state and at finite temperatures were investigated. The Ni-Fe-Al system 
is not only of high technological interest, but exhibits also rich physics,
e.g., a delicate interplay between structure and magnetism over a wide
composition range and substantial electronic correlations which is challenging
for modern electronic structure methods. The new Stuttgart ab-initio 
mixed-basis pseudopotential code in the generalized gradient approximation (GGA)
was used to determine the energetics in the ground state. Therewith, in 
combination with the cluster expansion (CE) method a representation of the 
energy landscape at $T$=0 over the whole Gibbs triangle was elaborated. At
finite temperatures, the cluster variation method (CVM) in tetrahedron
approximation was employed in order to calculate the ab-initio ternary phase 
diagram on the bcc and fcc lattice. Thereby, a miscibility gap in the ternary
B2 phase was theoretically verified.
\end{abstract}

\end{frontmatter}

% main text
\section{Introduction}\label{intro}
\subsection{Theoretical background}
From a theoretical point of view Materials Science can be understood as the study
of the interplay between realistic many-body systems. Although a lot of simplified
theoretical 
models to describe the physical properties of real materials work on a meso- or
macroscopic level with coarse-grained variables, these models often exhibit
badly defined materials parameters which stem from the true many-body 
problem. Thus, a microscopic ab-inito theory to model real materials systems
is not only appreciated to shed light on the definition of such parameters,
but also to provide a valuable tool to identify the important interactions and
physical processes on an atomistic level that determine the macroscopic 
behavior. In recent years it became possible by the invention of effective
approximations and the overwhelming increase of computer power to get a grip on
the highly demanding calculational schemes of such microscopic modellings from
first principles \cite{Tur94}.\\
The ab-initio density functional theory (DFT) in the local-spin-density 
approxmiation (LSDA) and in the generalized-gradient approximation (GGA) is 
nowadays a widely accepted highly efficient method to describe the electronic 
structure of weakly correlated materials. In DFT the electronic many-particle 
problem in the ground state is exactly mapped on an effective single-particle 
formalism, whereby the many-electron wavefunction is replaced by the electronic 
charge density $n({\bf r})$ as the characterising quantity of the system
\cite{Hoh64,Koh65}. In actual calculations only a small part of the total 
energy of the system, the exchange-correlation energy $E_{xc}[n]$, has to be 
approximated.\\
Due to the fact that in order to reveal the energetics of a given material 
system for arbitrary concentrations of the involved components a sole DFT 
description is still too costly, a suitable coarse-graining of the method 
has to be accomplished. This can be achieved in a well-defined way by the 
cluster expansion (CE) \cite{San84}. By this method, any function of the 
configuration on a given parent lattice can be expanded into cluster 
functions, whereby a cluster on the lattice is uniquely defined by the lattice
points that are included in such a geometrical object. The CE
coefficients are configuration independent. Through the structure inversion
method (SIM) \cite{Con83}, a direct link between the CE and the ab-initio 
electron theory can be established by fitting the coefficients of the CE 
(terminated at a maximum cluster) to the ab-inito calculated data for the physical
property which is to be cluster expanded. Because of the decoupling of expansion 
coefficients and occupations on the parent lattice, a CE established in such a way
can be highly efficient in obtaining values for the physical quantity in question 
for arbitrary lattice configurations and arbitrary concentrations of the 
components. By construction, these new values should in principle have the same 
accuracy as those explicitly computed via the underlying electronic structure 
method.\\
For the ab-initio description of finite temperature properties a modelling of
the free energy of the material system has to be achieved. This can be done 
by including entropy in the cluster theory in the form of the cluster 
variation method (CVM) \cite{Kik51,Mor57,San84,Fin94}. In the CVM one can 
write down a functional for the free energy where the internal energy is 
expressed by a CE with the cluster functions replaced by cluster correlation 
functions. The configurational entropy is also easily represented by the
latter \cite{San84}. By minimizing the free energy functional with respect
to the cluster correlation functions (or the cluster occupation probabilities
which can also serve as natural variables in this problem) a meaningful 
approximation to the true free energy of the system can be obtained. The 
approximation is of course given by the limited number of clusters one can 
consider in the practical cluster theory. In the limit of the inclusion of all
possible clusters, the CE and the CVM become mathematically exact on a given 
parent lattice.
\subsection{The Ni-Fe-Al system}
The present first-principles investigation was carried out for the ternary 
intermetallic system Ni-Fe-Al. Most of this type of modelling was done in the
past for binary systems, only few such ab-initio assessments can be found
in literature for ternary or even higher multicomponent materials systems. This
is in contrast to the technological development of new materials, where in most
cases more than two components are necessary to attain desired materials
properties. As an example for this serve alloys which are based on Ti, Fe 
and Ni aluminides \cite{Sau95}. Structural alloys derived from these 
intermetallic phases are promising candidates for high performance materials, 
because
they show high strength at all temperatures, low density and high chemical
resistance. The main drawback of these aluminides is their brittleness, in
particular at low temperatures. To overcome this drawback and also to further
improve the advantageous materials properties, either substitution of atoms 
by impurities or addition of other phases to the given one have to be
considered. For instance, small additions of Boron to Ni$_3$Al have a 
ductilization effect \cite{Liu95}. Concerning NiAl, it was found that minor
amounts of Fe additions also enhance the ductility \cite{Dar92}. Furthermore, 
as shown by Letzig \textsl{et al.} \cite{Let99} the complete Ni-Fe-Al system 
offers various alloying possibilities for softening NiAl and Ni$_3$Al, too.\\ 
But a theoretical investigation of this ternary system may be motivated not only 
from a purely technological point of view. Besides the general 
need for consolidation of the first-principles alloy theory for multicomponent
systems, i.e., higher than binary, the Ni-Fe-Al system exhibits also rich
physics from a very fundamental point of view. The system incorporates two
transition metals, Ni and Fe, with partially filled $3d$ states which can give
rise to localized electronic behavior, and the simple metal Al with itinerant
$s$ and $p$ states. Moreover, Ni and Fe are ferromagnetic in their elemental
crystal ground state, whereby the structure differs, i.e., Ni crystallizes on 
a fcc and Fe on a bcc lattice. Nonmagnetic Al is again stable in a fcc 
structure, but with a much larger lattice constant compared to fcc-Ni, since
the atomic radius of Al is 15\% bigger than the atomic radius
of Ni (Ni and Fe atoms have nearly identical atomic radii in the solid state).
Therefore, depending 
on the concentration of the three constituents, a strongly varying behavior in
the electronic structure, magnetism and the alloy structure is
expected. Already well studied are the binary subsystems, i.e., Ni-Fe, Ni-Al
and Fe-Al (see \cite{Swa91,Nas91,Kat93} and references therein). The Ni-Fe 
system, with the famous Invar region \cite{Ent93} around Ni$_{35}$Fe$_{65}$, 
shows a very strong influence of magnetism on the phase stability. In contrast, 
in Ni-Al this influence is nearly negligible. Lying between these limiting 
cases, Fe-Al shows a very delicate interplay between structure and magnetism 
\cite{Kat93,Min86,Lec02,Lec04}. In addition, the Al-rich aluminide systems exhibit
low ordering energies that give rise to stable crystal structures with 
low symmetries, whereas for the transition-metal-rich regions only structures
on the bcc and fcc lattice are stable.\\
Only very few theoretical approaches to model Ni-Fe-Al are 
known from literature (e.g., \cite{Shi97,Eno91}). Especially, there exists no 
complete theoretical 
description for the global system, also not with empirical parameters. There
are experimental investigations (for instance, \cite{Riv80,Bud92} and references 
therein), but the overall understanding of this system, in particular on an
atomistic level, is rather scarce. Hence the present work is dedicated to provide
an introduction to an ab-initio modelling of Ni-Fe-Al by revealing some insight in
the interesting physics of this system.
\section{Calculational procedure}
\subsection{DFT calculations}
To elucidate the ground state properties of Ni-Fe-Al, as a first step DFT 
calculations with our ab-initio mixed-basis pseudopotential (MBPP) code  
\cite{MBPP} were performed for chosen reference solid compounds. In the
MBPP code, norm-conserving pseudopotentials are used and the basis consists of 
plane waves and a few additional localized 
functions per atom. For $E_{xc}$ we used the GGA given by Perdew, Burke and
Ernzerhof (PBE) \cite{Per96}. In the whole modelling only structures on the
bcc and fcc lattice were taken into account (with a small exception for Al-rich
Ni-Al at $T$=0, cf. section \ref{binsub}), thereby restricting the Ni-Fe-Al
system onto these cubic parent lattices. As other lattice types become important 
only in the technologically less interesting Al-rich region of the ternary phase 
diagram \cite{Riv80,Bud92}, this restriction cuts off only a small part of
common general interest. To investigate the influence of collinear magnetism we
performed both spin-unpolarized and spin-polarized calculations for 
each considered crystal structure.\\
In the MBPP calculations, the numbers of k-points in the irreducible wedge of 
the 1. Brillouin zone where 408, 440 and 280 for fcc-Ni, bcc-Fe and fcc-Al,
respectively, in order to ensure an equivalent k-point grid for the different
structures. With the same intention, the k-point grid for all the treated
compounds was chosen in such a way that the density of sampled k-points was
the same for each structure. The cut-off energy for the plane waves
was generally set to 24 Ryd, and throughout the calculations a Gaussian smearing
of 0.05 eV was employed. With this choice for the convergence parameters the
relevant formation energy $E_f$, defined for a given structure 
$\vc\sigma$ as
\begin{equation}
E_f(\vc\sigma)=E_{tot}(\vc\sigma)
-c_{\sub{Ni}}E_{tot}^{\sub{(fm)}}(\mbox{fcc-Ni})
-c_{\sub{Fe}}E_{tot}^{\sub{(fm)}}(\mbox{bcc-Fe})
-c_{\sub{Al}}E_{tot}^{\sub{(nm)}}(\mbox{fcc-Al}),\label{eform}
\end{equation}
with the concentration $c_i$ of the species and the total energy $E_{tot}$,
whereby the superscripts 'nm'/'fm' indicate the nonmagnetic/ferromagnetic state,
can be converged to 0.05 meV/atom.
\subsection{Cluster expansions\label{cluexp}}
\begin{figure} % Fig 1: clusters
\epsfig{file=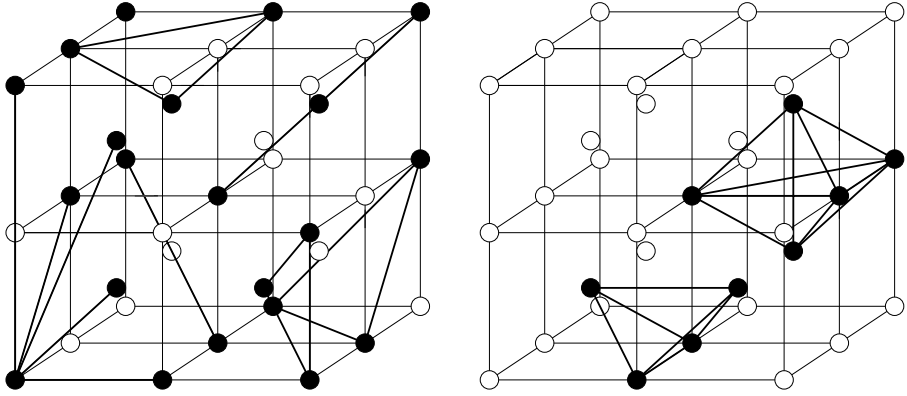,width=12cm}
\caption{Cluster used for the pentahedron approximation on the bcc 
lattice.\label{figure1}}
\end{figure}

From the calculated formation energies and magnetic moments, a set of CEs was 
constructed via the SIM. In practise, a CE for the function $f$ of the 
configuration $\vc\sigma$ in an $M$ component system on a given parent 
lattice can generally be written as \cite{San84}
\begin{equation}
f(\vc\sigma)=f_0+\sum_{\alpha,m}^{\gamma,M-1}f_{\alpha m}\phi_{\alpha m}
(\vc{\sigma}_{\alpha})\quad,\label{ce}
\end{equation}
where $\phi_{\alpha m}$ is the cluster function for cluster $\alpha$ and
degree $m$, $f_{\alpha m}$ is the cluster expansion coefficient and $\gamma$ 
is a maximum cluster. The quantity $f_0$ designates the empty cluster. The cluster
function is suitably chosen to be a function of spin variables $\sigma$. For the 
considered ternary case values of 1, -1 and 0 where attributed to these occupation
variables assigned to Ni, Fe and Al. As the complete ternary basis for the cluster
functions $\phi_{\alpha m}$, we used $\Theta_1$=$\{1,\sigma,\sigma^2\}$, and for 
comparison also the orthonormal Chebychev basis \cite{San84,Wol94}, below named 
$\Theta_2$. The clusters $\alpha$ belong to a chosen set with a maximum cluster 
$\gamma$, where $\alpha$ does not necessarily have to be a subcluster of $\gamma$.
The expansion coefficient $f_{\alpha m}$ will be named $K_{\alpha m}$ in the case
of the formation energy. Note that generally any CE coefficient can be understood 
as an effective cluster interaction (ECI) which includes contributions from 
physical interactions within and outside the cluster range \cite{Fon94}.\\ 
In our work, CEs for the bcc and fcc lattice were constructed.
Two cluster sets were used. First, we extracted for both parent lattices a 
traditional tetrahedron CE (T-CE) (see \cite{Ind01} and references therein) from 
the ab-initio calculated data for the physical quantities in question, i.e., 
formation energy and magnetic moment, of 
selected reference configurations. In this T-CE, all used clusters are subclusters
of the minimum tetrahedron on the given lattice. On the bcc lattice this includes 
5 clusters, namely the point, the nearest-neighbor (NN) pair, the 
next-nearest-neighbor (NNN) pair, the triangle with two NN pairs and one NNN 
pair and the irregular tetrahedron itself with two NN pairs and two NNN pairs. For 
the fcc lattice the T-CE corresponds to 4 clusters, i.e., the point, the NN pair, 
the triangle with three NN pairs and the regular tetrahedron with four NN 
pairs. Hence, whereas the bcc T-CE incorporates NNN correlations, 
the fcc T-CE includes only NN correlations. Due to the completeness 
requirement, there are 21 cluster functions on the bcc lattice and 15 on the 
fcc lattice in the ternary case for the tetrahedron approximation. The same 
numbers of ECIs were determined by the SIM via matrix inversion, i.e., the 
number of reference structures was equivalent to the number of ECIs.\\
As we were mainly interested in the ternary B2 phase of the Ni-Fe-Al system, 
we developed in addition a higher CE for the bcc lattice with 13 clusters, 
including the point, pairs up to the 6th NN, four triplets, the irregular
tetrahedron and a pentahedron (s. Fig. \ref{figure1}). It follows that 
there are 65 different cluster functions, and associated therewith, of 
course, the same number of ECIs, for the ternary system. In that
case the SIM was performed via a least-mean-squares fit of 127 ab-initio 
calculated formation energies and magnetic moments, out of which 75 belonged to 
binary and 49 to ternary compounds. We will name this CE, according to the maximum
body  cluster, 'pentahedron CE (P-CE)' but note that in contrast to the T-CE, this
does not mean that all and only the subclusters of this pentahedron are included.\\
With these cluster sets, the formation energy and the magnetic moment were 
expanded in different schemes. First, the spin-unpolarized and spin-polarized 
formation energy were expanded for both parent lattice types with allowing 
only for global volume relaxations but no structural relaxations of the reference 
structures, in order to reveal the influence of magnetism in the fixed geometries. 
Thereby, as an approximation we allowed only for ferromagnetic (fm) alignment of 
the local moments which possibly develop in the spin-polarized MBPP calculations.
This can be validated by the experimental observation that all stable cubic 
ordered compounds in the Ni-Fe-Al system are ferromagnetic in their respective 
ground state when showing magnetic behavior. The only critical case is 
stoichiometric FeAl, where DFT in LSDA/GGA predicts a ferromagnetic ground state 
\cite{Min86}, whereas experimentally no net magnetic moment at very low 
temperatures is detected \cite{Cas73}. Throughout the rest of the paper the 
nomenclature $E^{\sub{(fm)}}_f$ will be used for the cluster-expanded formation 
energies of structures with potentially ferromagnetic order, i.e., the magnetic 
energy is implicitly included in that CE.\\
Additionally, we checked for the influence of ``local structural relaxations'' on 
the 
bcc lattice by cluster-expanding also the spin-polarized formation energy for the 
structurally relaxed reference structures. During a ``local structural relaxation''
all atoms are relaxed but the unit cell is kept cubic, i.e., no shape relaxation
of the supercell is performed. For instance, among the various considered
configurations in the supercell there are such which would lead to a tetragonal
distortion of the supercell if we allowed for a c/a relaxation. To keep the cubic 
shape of the supercell during structural relaxation seems to be meaningful for a 
modelling with multiple parent lattices, as employed here. Within a 
``full structural relaxation'' including shape relaxation, the connection between 
the
reference structure and its supposed parent lattice may get lost more easily. An
example is the case of tetragonal L1$_0$-NiAl on the fcc parent lattice, which
transforms into B2-NiAl on the bcc parent lattice when relaxing $c/a$ \cite{Wol00}.
The described way of including the structural relaxations was performed only in 
the more sophisticated pentahedron approximation. Throughout the rest of the paper,
the resulting CE is correspondingly named rP-CE.\\
The mean error in meV/atom for the ECIs from the least-mean-squares fitted SIM was 
13.4, 11.1 and 12.0 for the spin-unpolarized P-CE, the spin-polarized P-CE and the
rP-CE (spin-polarized) in the $\Theta_1$ basis, respectively. Hence, the latter 
CEs should be well enough converged to account for a meaningful description of the
energetics in the given approximations.\\
In order to investigate the relevant energy landscape of Ni-Fe-Al by the 
constructed ternary CEs, we developed a simple algorithm to determine homogeneous 
lowest-energy structures for each given ternary composition. In this algorithm, a 
random starting configuration is successively modified by site exchange of, 
respectively, two atoms in a periodically continued supercell. After each 
such Monte-Carlo sweep, the energy of the new structure is calculated and the 
structure will be accepted if the energy is lower than the one of the old 
structure. For the supercells, sizes of 54 and 108 atoms were used in the 
calculations. The minimum formation energy is usually reached in some thousand 
sweeps and the corresponding optimized homogeneous structure may be extracted. 
Please note that the so obtained homogeneous lowest-energy structures for 
arbitrary compositions are not necessarily identical to the true ground state 
structures which are often heterogeneous mixtures of various phases.
\subsection{CVM calculations}
For the calculation of the ternary incoherent phase diagram on the bcc and fcc
lattice at finite temperatures, the CVM in tetrahedron approximation was employed
for the internal energy and the configurational entropy. The ab-initio ECIs 
$\{K^{\mbox{\tiny{fm}}}_{\alpha m}\}$ obtained from the spin-polarized MBPP 
calculations were used for the representation of the internal energy. By confining
to ferromagnetic order in the MBPP calculations the effect of thermal magnetic
excitations on the finite temperature phase diagram is neglected. The actual CVM 
calculations were performed in a grand canonical ensemble with the set of 
effective chemical potentials 
$\{\tilde{\mu}_{Ni},\tilde{\mu}_{Fe},\tilde{\mu}_{Al}\}$ and the effective grand 
potential $\Omega(T,V,\{\tilde{\mu}_i\})$, whereby the CVM equations were 
solved by the Natural Iteration Method (NIM) \cite{Kik74}. This whole procedure
is described in detail elsewhere \cite{Col93}. Although vacancies are in 
principle important defects in phase diagram calculations of intermetallic 
aluminide systems \cite{Lec01}, they were not taken into account in this 
investigation. In a coherent modelling, this would call for a quaternary 
description with the vacancy as a fourth component with chemical potential zero.
Also, vibrational effects on the thermodynamics were neglected.
\section{Results}
\subsection{Zero temperature}
\subsubsection{General remarks\label{genmark}}
In the Tabs. \ref{table1} and \ref{table2} the structural data for the structures
on the bcc and fcc parent lattice used for the development of the respective CEs 
in tetrahedron approximation, i.e., T-CEs, are shown. All structures correspond to
a specific tetrahedron occupation on the respective parent lattice \cite{Ind01}, 
hence these compounds have a small unit cell and order in NN and NNN neighbor 
distances. The first line for each compound in Tabs. \ref{table1} and 
\ref{table2} corresponds to a spin-unpolarized calculation, the second line to
a spin-polarized one with possible fm order. The lattice constants are generally 
increased and the bulk moduli are decreased by ferromagnetism. The compounds with 
a '*' are confirmed ground states in the Ni-Fe-Al system \cite{Bud92}. Thus, none 
of the highly ordered fully ternary structures are stable within the temperature
regime investigated in experiment. From our MBPP calculations the necessary 
condition for phase stability, i.e., a negative formation energy, is fulfilled by 
all these structures, at least when taking into account ferromagnetism. However, 
by checking the possible tie line constructions \cite{Wat98} within the given set 
of reference structures, we can confirm the experimental observation from the 
$T$=0 viewpoint, although the case of NiFeAl$_2$ is delicate (more on this 
stoichiometry in section \ref{tersys}).\\
The CE of the possibly ferromagnetic compounds accounts implicitly for the 
magnetic degrees of freedom. Thereby it is assumed that these degrees of 
freedom are strongly coupled to the chemical degrees of freedom which are 
treated explicitly in Eq. (\ref{ce}). Furthermore, as already mentioned, 
other spin orderings were not allowed, and concerning the energy it was assumed 
that the energies of such possible non-fm configurations can be approximated by 
the fm energy contribution. Both approximations might be critical in some regions
of the Ni-Fe-Al phase diagram, e.g., the Invar region in Ni-Fe \cite{Sch99}, 
but should be reliable for a qualitative inspection of the general trends 
in the whole system.

The ECIs for the CEs of the formation energy in tetrahedron approximation are
shown in Tab. \ref{table3} for the nonorthogonal $\Theta_1$ and the 
orthonormal $\Theta_2$ basis. Although both sets of ECIs stem from an exact SIM 
with the same reference
structures, the decay of the absolute value of the ECIs is generally more clearly
visible within the $\Theta_2$ basis. However, in practical convergence tests via 
the prediction of energies of new structures not included in the SIM, no 
appreciable differences between the two basis sets were observed (for a further
discussion on this matter see, e.g., \cite{Wol94}).\\
The discussion of the ECIs in a ternary system is not as straightforward as
in the binary case, because due to the completeness relation of the CE there are 
now multiple interactions associated with each geometrical cluster. A first
estimation of the ordering/segregation behavior is, for instance, not easily 
connected to the sign of the effective pair interactions (EPI) from the
common CE \cite{Wol94}. Thus, in order to illustrate this behavior it is more 
appropriate to map the obtained ternary EPIs onto effective quasibinary pair 
interactions $W^{(n)}_{NiFe}$, $W^{(n)}_{NiAl}$ and $W^{(n)}_{FeAl}$ \cite{Wol94},
where $n$ denotes the near neighbor distance, that again yield a first 
information about the ordering/segregation behavior of the system via their 
sign. Thereby, positive (negative) values describe ordering 
(segregation) behavior. The values for these quasibinary EPIs, obtained from a 
mapping of the 
ternary EPIs from Tab. \ref{table3}, are shown in Tab. \ref{table4}. It is seen 
that the $W^{(n)}$ between the transition-metal atoms and the Al 
atoms are always of an ordering type, in the case of $W^{(n)}_{FeAl}$ subtly 
influenced by magnetism. In contrast, the NiFe interactions on the bcc lattice 
remain negative in NN distance even when ferromagnetism (very important in the 
binary Ni-Fe system, see Tab. \ref{table1}) is taken into account. Only on the 
fcc lattice with ferromagnetism there is a indication for weak ordering 
tendencies. Although the NN pair interactions seem to be dominant in Ni-Fe-Al, of 
course, the effect of the multiplet and also of the higher pair interactions 
beyond NNN distance is neglected in the simple set $\{W^{1,2}_{AB}\}$.
\subsubsection{Binary subsystems\label{binsub}}
\begin{figure} % Fig 2: binaries
\psfrag{Ni-Fe}{Ni-Fe}\psfrag{Ni-Al}{Ni-Al}\psfrag{Fe-Al}{Fe-Al}
\psfrag{cni}{c$_{\sub{Ni}}$}\psfrag{cfe}{c$_{\sub{Fe}}$}
\psfrag{Emevatom}{$E^{\sub{(fm)}}_f$ [meV/atom]}
\psfrag{Mmubatom}{$M$ [$\mub$/atom]}
\epsfig{file=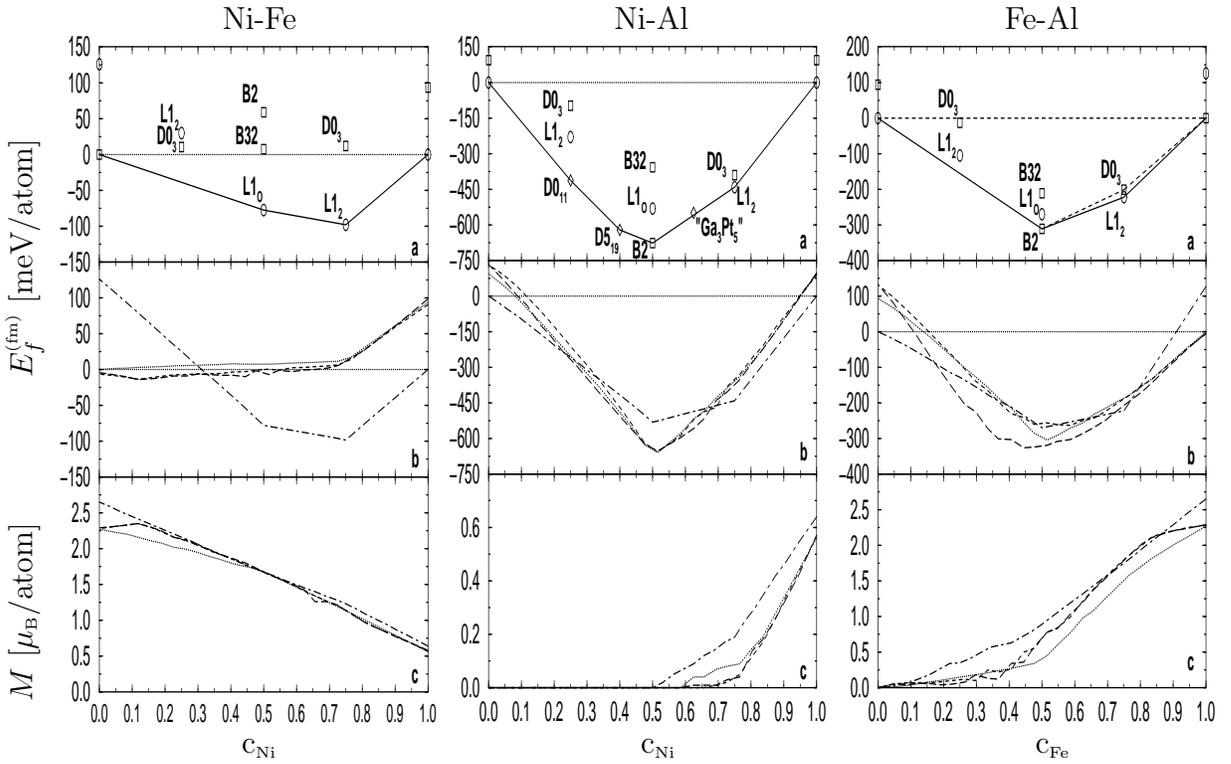,width=16cm}
\caption{Formation energy $E^{\sub{(fm)}}_f$ and magnetic moment $M$
in the binary systems Ni-Fe, Ni-Al and Fe-Al. (a) $E^{\sub{(fm)}}_f$ according 
to the spin-polarized MBPP calculations. (b) $E^{\sub{(fm)}}_f$ according to 
the constructed CEs. (c) $M$ according to the constructed CEs. In (b) and (c) 
the dotted lines correspond to the bcc T-CE, dashed lines to the 
bcc P-CE, long-dashed lines to the bcc rP-CE and dotted-dashed lines to the 
fcc T-CE.\label{figure2}}
\end{figure}

For the whole composition range, Fig. \ref{figure2} shows the ferromagnetic 
formation energy and magnetic moment for homogeneous periodic configurations of 
the binary subsystems of Ni-Fe-Al at zero temperature. In Fig. \ref{figure2}a
the MBPP formation energies for the binary ''canonical'' tetrahedron-representable
ordered structures from Tab. \ref{table1} and \ref{table2} are depicted. 
Additionally, for Ni-Al the formation energies of the noncubic 
experimentally stable structures D0$_{11}$-NiAl$_3$, D5$_{19}$-Ni$_2$Al$_3$
and ``Ga$_3$Pt$_5$''-Ni$_5$Al$_3$ are also included. The latter structures
were structurally relaxed within the MBPP code, whereas the canonical structures 
are cubic structures (apart from the ones with L1$_0$ symmetry) for which no 
structural 
relaxation degrees of freedom exist. The L1$_0$ structures were restricted to the 
fcc parent lattice (see section \ref{cluexp}). The ground state structures
according to the MBPP calculations are connected with straight lines in Fig. 
\ref{figure2}a. For Ni-Fe, L1$_0$-NiFe was also theoretically verified as a 
stable ground state structure within the investigated set, although 
experimentally this is still questionable \cite{Swa91}. In the Ni-Al system the 
correct ordered ground states according to experiment are obtained, except for 
``Ga$_3$Pt$_5$''-Ni$_5$Al$_3$ which we were not able to verify as a stable
ground state structure (also when including structural relaxations). Problematic 
seems to be the case for Fe-Al, as in PBE-GGA
the important D0$_3$-Fe$_3$Al structure was not identified as the stable
ground state at Fe$_{75}$Al$_{25}$, but L1$_2$-Fe$_3$Al. This seems to be due to 
the subtle electronic structure in this region of Fe-Al, where simple
approximate exchange-correlation functionals, i.e., LSDA and GGA, might be
not appropriate. For a discussion of this peculiarity see \cite{Lec02,Lec04}.\\
The zero temperature description according to the CEs constructed from the
reference structures including spin-polarization, shown in Fig. \ref{figure2}b,
yields important additional information. The lines in Fig. \ref{figure2}b
correspond to the curves of lowest formation energy for the homogeneous structures
obtained from the Metropolis-inspired algorithm described at the end of 
section \ref{cluexp}. To obtain these curves, a homogeneous structure optimization
for 26 concentrations in the interval [0,1] was performed with this algorithm, and
the minimum formation energy was extracted, respectively. 
First, the formation energy in Fig. \ref{figure2}b shows for
Ni-Fe major energy differences between phases on the bcc and the fcc parent 
lattice. The 
energetical crossover lies just around the Invar region. One can also see the 
important change in the sign of $E^{\sub{(fm)}}_f$ for structures on the bcc 
lattice from positive to negative when applying the pentahedron CEs, especially
for the Fe-rich system. However, the corresponding phases are metastable because 
their formation energies lie above the connecting line between the ground state
structures bcc-Fe, L1$_0$-NiFe, L1$_2$-Ni$_3$Fe and fcc-Ni. The magnetic moments of
these metastable phases are substantially enhanced, as seen by the maximum in the 
corresponding curves in Fig. \ref{figure2}c. Turning over to Ni-Al, the large 
negative formation energy curves in this system exhibit a strong symmetry with 
respect to the equiatomic composition. These energies show a similar qualitative 
compositional dependence on both cubic parent lattices, whereby the bcc lattice is
favoured around the equiatomic composition. The D0$_{11}$-NiAl$_3$ phase lies 
obviously in the bcc/fcc transition region, the same applies for the Ni$_5$Al$_3$ 
stoichiometry (see also \cite{Slu92}). Although our calculations do not stabilize 
``Ga$_3$Pt$_5$''-Ni$_5$Al$_3$, the large relaxation energy around this composition
yields a hint for a possible martensitic transition. In general, the magnetic 
energy does not seem to be essential, although in the Ni-rich region it has some 
influence which can be relevant for the phase stability \cite{Wol99}. The 
increase of the magnetic moment with increasing Ni content is also clearly seen in
Fig. \ref{figure2}c. Finally, it is apparent from the figure that the energetical 
competition between the bcc and fcc parent lattice becomes 
most delicate for the case of Fe-Al. The formation energy curves for both lattice 
types are very similar over a wide composition range. Also, the relaxation energy 
appears to be crucial in determining the correct phase stability. Especially 
around the equiatomic composition, the differences between the various 
approximations are significant. In contrast to Ni-Al, the magnetic moment persists
in the Al-rich region of Fe-Al and may have relevant influence on the already 
sensitive energy landscape. As already stated, the existence of a net magnetic 
moment at equiatomic FeAl is experimentally not verified, but the sensitivity to
magnetic degrees of freedom is nevertheless experimentally confirmed 
\cite{Bog98}, also for larger Al content \cite{Lue01}.
\subsubsection{Ternary system\label{tersys}}
\begin{figure} % Fig 3: ternaries
\parbox{5cm}{a\\
\epsfig{file=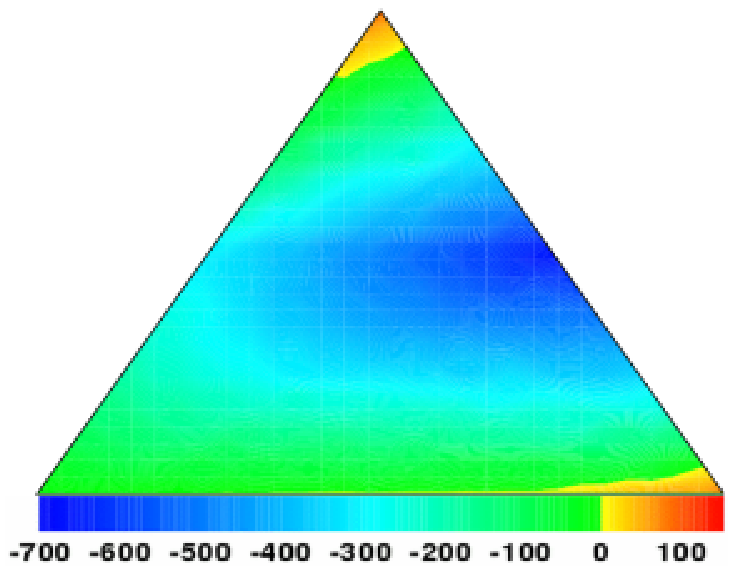,width=7cm}}
\parbox{3.5cm}{\hfill}
\parbox{5cm}{b\\
\epsfig{file=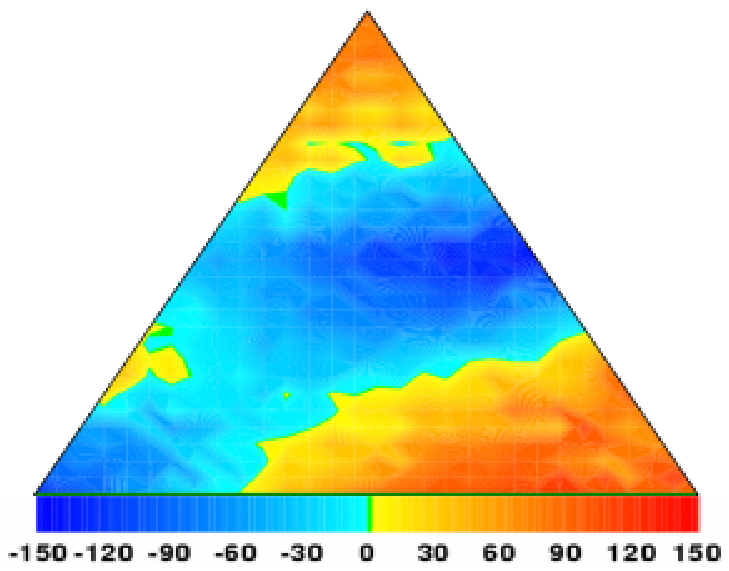,width=7cm}}\\[0.2cm]
\parbox{5cm}{c\\
\epsfig{file=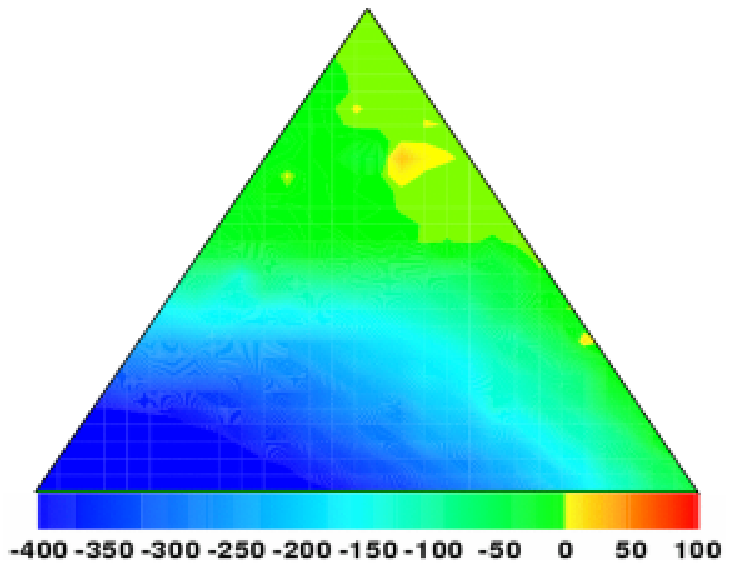,width=7cm}}
\parbox{3.5cm}{\hfill}
\parbox{5cm}{d\\
\epsfig{file=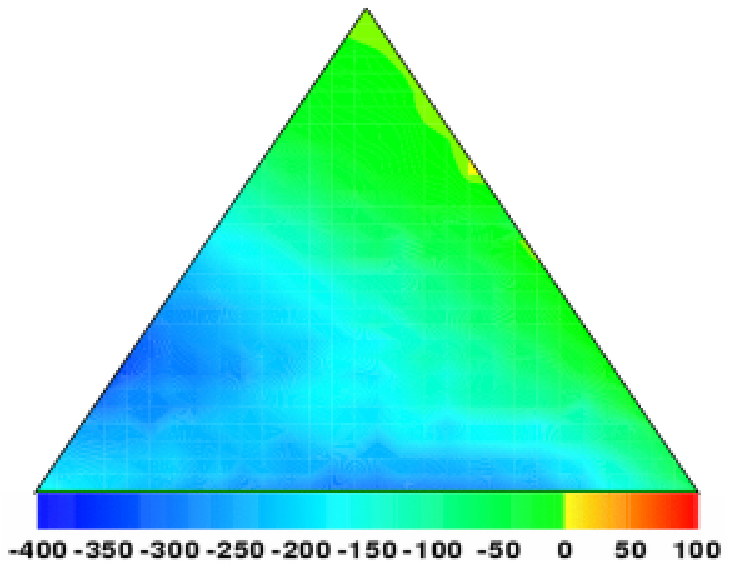,width=7cm}}\\[0.2cm]
\parbox{5cm}{e\\
\epsfig{file=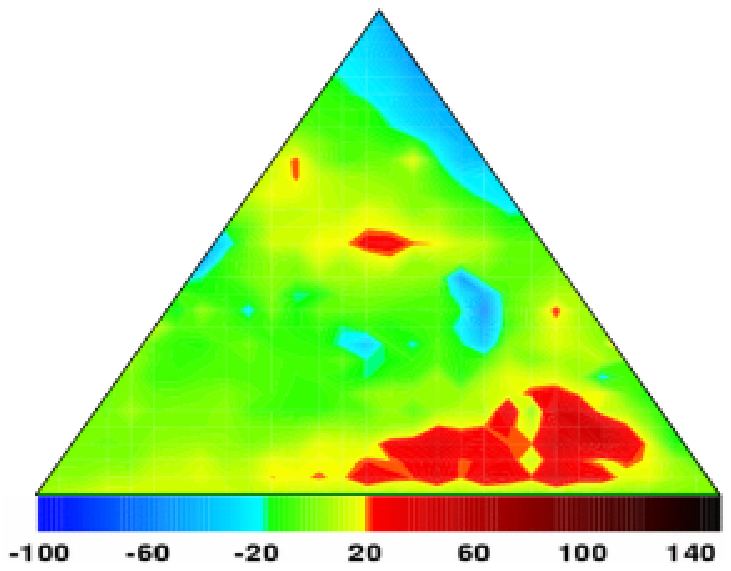,width=7cm}}
\parbox{3.5cm}{\hfill}
\parbox{5cm}{f\\
\epsfig{file=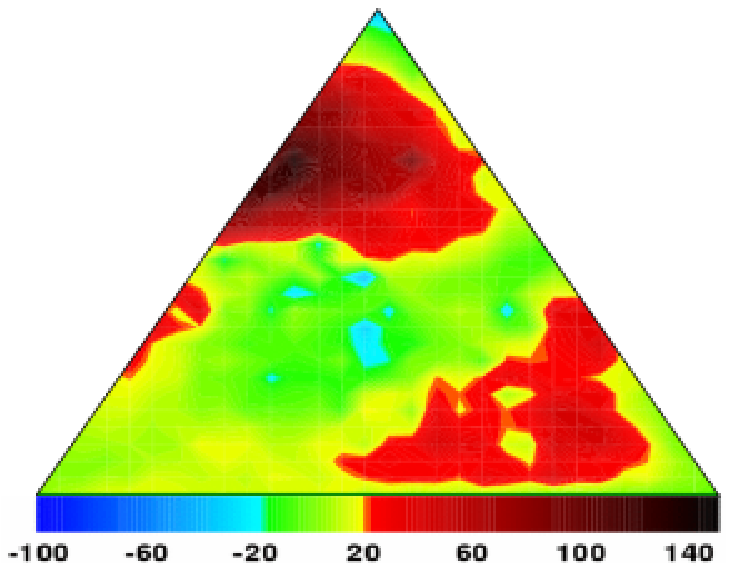,width=7cm}}
\caption{Ternary formation energies $E^{\sub{(fm)}}_f$ and differences 
$\Delta E_f$ (in meV/atom). The Gibbs triangle is oriented with the Al corner at
the top, Fe corner on the left and Ni corner on the right. All energies in the
series (a)-(d) belong to the CEs in tetrahedron approximation. 
(a) $E^{\sub{(fm)}}_f$ on 
the bcc lattice. (b) $\Delta E^{\sub{(fm)}}_f(\mbox{bcc-fcc})$ 
between the formation energy on the bcc and fcc lattice. (c)  
$\Delta E^{\sub{(fm)}-\sub{(nm)}}_f(\mbox{bcc})$ between the fm and nm formation
energy on the bcc lattice. (d) 
$\Delta E^{\sub{(fm)}-\sub{(nm)}}_f(\mbox{fcc})$ between the fm and nm formation
energy on the fcc lattice. (e) Difference between $E^{\sub{(fm)}}_f$(bcc) in
tetrahedron and pentahedron approximation. (f) Difference between 
$E^{\sub{(fm)}}_f$(bcc) in tetrahedron and relaxed pentahedron (rP-CE) 
approximation.
\label{figure3}}
\end{figure}

\begin{figure} % Fig 4: NiFeAl_2
\psfrag{cni}{c$_{\sub{Ni}}$}\psfrag{cFe}{c$_{\sub{Fe}}$}
\psfrag{Eformmevatom}{$E^{\sub{(fm)}}_f$ [meV/atom]}
\psfrag{Mmubatom}{$M$ [$\mub$/atom]}
\epsfig{file=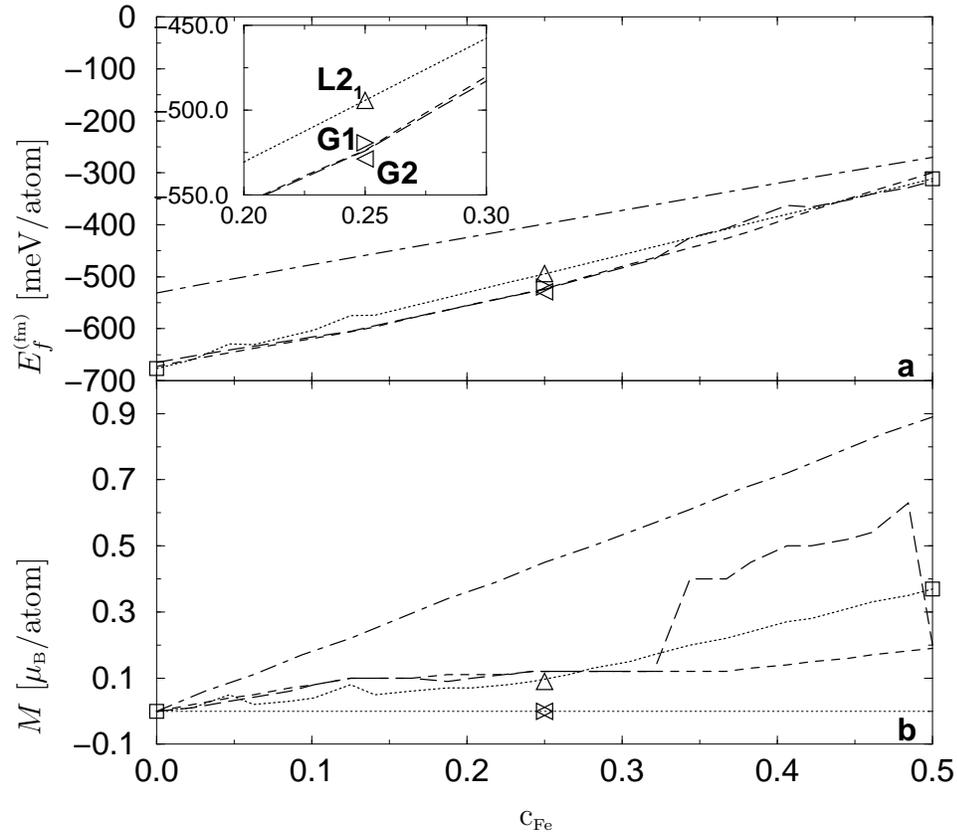,width=12cm}
\caption{Formation energy $E^{\sub{(fm)}}_f$ (a) and magnetic moment $M$ (b) along
the NiAl$-$FeAl line in the Ni-Fe-Al Gibbs triangle. Symbols denote the MBPP
values for the ordered structures. Lines are according to Fig. \ref{figure2}.
\label{figure4}}
\end{figure}

\begin{figure} % Fig 5: crystal structures for NiFeAl_2
\psfrag{a}{a}\psfrag{b}{b}\psfrag{c}{c}
\epsfig{file=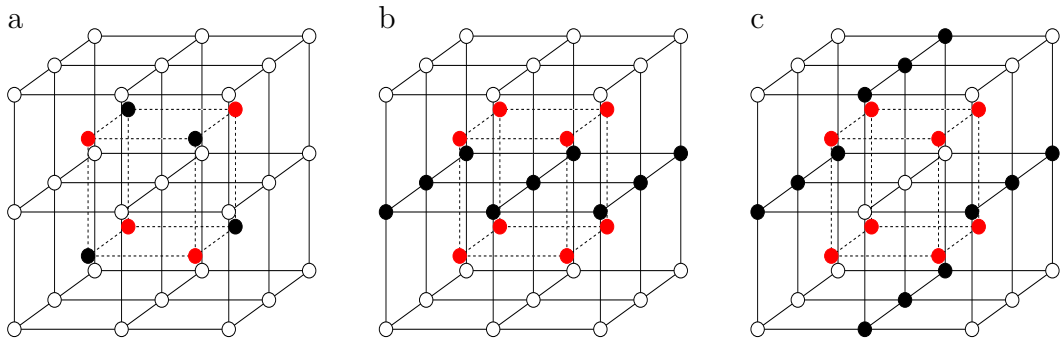,width=14cm}
\caption{Important ordered structures at NiFeAl$_2$. 
(a) L2$_1$, (b) G1 and (c) G2. Open circles denote the Ni atoms, black circles
the Fe atoms and (red/grey) circles the Al atoms.\label{figure5}}
\end{figure}

Turning now to the multicomponent system, Fig. \ref{figure3} shows lowest 
formation 
energy landscapes for ternary Ni-Fe-Al in the Gibbs triangle representation. These
landscapes were computed again via our algorithm to find the minimum formation 
energy utilizing the respective CEs. Therefore, we introduced a 2-dimensional mesh
of 338(254) grid points for the bcc(fcc) parent lattice inside the Gibbs triangle.
At each of these grid points we searched for the homogeneous phase with the
lowest formation energy within a given periodically continued supercell. Please
recall that in this kind of calculations we again did not perform any tie line 
construction to account for heterogeneous phases, which would result in a true 
ground-state phase diagram.\\
On both cubic parent lattices, the respective minimum formation energy is nearly 
exclusively negative inside the Gibbs triangle. Only close to the binaries or to 
the pure elements, $E^{\sub{(fm)}}_f$ reaches positive values in selected cases. 
The topology of the formation energy landscape appears to be very similar on the 
two parent lattices. Generally, $E^{\sub{(fm)}}_f$ exhibits the largest negative
values around the region with 50\% Al, with a deep valley leading towards the 
Ni$_{50}$Al$_{50}$ composition (Fig. \ref{figure3}a). Important information 
concerning the relative stability of the two cubic parent lattices is provided by 
Fig. \ref{figure3}b which displays the difference between the obtained formation
energies for phases on the bcc and the fcc lattice. Clearly seen is that the 
phases on the fcc lattice dominates the Ni-rich and the Al-rich region of 
Ni-Fe-Al. The range where the homogeneous phases on the bcc lattice have lower 
formation energy than those on the fcc lattice resembles a belt shape, oriented 
along Fe-NiAl. Thereby, two stability centers may be identified. One is located at
NiAl and the other distinct one is centered at the Fe corner of the triangle. The 
bridge between these centers becomes very fragile around the NiFe$_2$Al 
composition, indicating a strong competition with phases on the fcc lattice in 
this region. This competition was already experimentally verified in early 
studies of the Ni-Fe-Al system (see \cite{Riv80} and references therein). The 
nearby problematic Fe$_3$Al section (see section \ref{binsub}) underlines the 
delicate 
energetics in this area. Regarding the influence of ferromagnetism on the 
formation energy (see Figs. \ref{figure3}c,d), it is obvious that the main magnetic
energy contributions arise for Fe-rich alloys. On the fcc lattice, the magnetic 
energy contribution exhibits maxima around NiFe and Fe$_3$Al, whereas on the bcc 
lattice one deep funnel centered at the Fe corner appears. In general, the 
behaviour of the cluster-expanded fm moment follows mainly the one of the magnetic
energy over the Gibbs triangle. Although the influence of magnetism is manifest
in the Fe-rich region, even for compositions without a dominant fraction of Fe the
magnetic energy gain for a wide range of alloys is still larger than 50 meV/atom. 
Hence, a neglect of this important energy term could lead to a qualitatively 
wrong picture of the Ni-Fe-Al system. Finally, we checked for the importance of 
higher cluster correlations on the bcc lattice. Fig. \ref{figure3}e shows the 
difference between $E^{\sub{(fm)}}_f$ in tetrahedron and in pentahedron 
approximation. Additionally, in Fig. \ref{figure3}f the relaxed pentahedron 
approximation (rP-CE) was utilized to calculate this difference. One can observe 
that the tetrahedron approximation is not too bad for the bcc stability region. 
Deviations of the order of $\pm$20 meV/atom are surely comparable to the 
underlying DFT-GGA errors. Significant impact of the  higher correlations appears 
for the Ni-rich, and by turning on the effect of structural relaxations, also for 
the Al-rich region. In both cases the higher cluster correlations tend to further 
lower the optimal formation energy. This effect is intuitively clear for the 
Al-rich region because of the different atomic sizes of the transition metal 
atoms and the Al atom. By structural relaxations, the system can reduce internal
stress by lifting the constraints for the atomic positions imposed by the 
unperturbed parent lattice. Generally, for structures with cubic symmetry the 
ordering energy is small in the Al-rich region. The energy gain for Ni-rich 
bcc-alloys by taking into account larger clusters might be due to the fact that 
the bcc parent lattice is not energetically favorable in this region. The larger
clusters stabilize structures with bigger unit cell, for which the locally defined
``bcc character'' diminishes. In other words, the stabilization of large 
unit cells against small unit cells indicates a low ordering energy on the given
parent lattice. Recall that Ni-rich Ni-Fe-Al belongs to the region where phases on
the fcc lattice are energetically favorable.\\
In order to quantify the predictive power of our ternary CEs, we present at the 
end of these $T$=0 considerations a ground state investigation for the 
NiFeAl$_2$ composition. Remember that our Metropolis-inspired algorithm determines
the lowest-energy ordered structure for a given composition. However, this 
structure is only stable if in the diagram of the formation energy vs. compostion
its formation energy lies below the tie line connecting the formation energies of
at least two neighbouring phases. Otherwise, a heterogeneous mixture of the 
corresponding phases is thermodynamically more stable than the homogeneous phase.
When we want to check for a ternary system the stability of the lowest-energy
ordered structure for a given composition, we in principle have to investigate the
position of its formation energy relative to the tie lines corresponding to all
possible neighbouring phases in all directions of the Gibbs triangle \cite{Wat98}.
Because this is very time-consuming, after checking the energetics of the 
binaries, i.e., Figs. \ref{figure2}a,b, we confined the detailed study to the 
NiAl$-$FeAl quasibinary. As from the latter check, due to the strongly negative 
formation energy of B2-NiAl and B2-FeAl, phases on the NiAl$-$FeAl quasibinary 
line appear to be the most promising competitors against an homogeneous phase at 
NiFeAl$_2$.
Note that the NiAl$-$FeAl line also serves as a boundary for the importance of 
ferromagnetism with increasing Al content (see Fig. \ref{figure3}c), i.e., 
ordering tendencies along this line are also interesting from that point of
view. From experiment, no ordered structure is known for NiFeAl$_2$. In Fig.
\ref{figure4}a the formation energies along NiAl$-$FeAl according to the different
CEs are plotted. First, the fcc T-CE yields a noticeably higher formation energy 
curve than the simple B2-NiAl$-$B2-FeAl tie line, i.e., the fcc parent lattice 
provides no promising candidates concerning an ordered ground state. Turning to 
the bcc lattice, the $E^{\sub{(fm)}}_f$ curve belonging to the T-CE is nearly 
identical to the straight B2-NiAl$-$B2-FeAl tie line, and the simple Heusler L2$_1$
structure (see Fig. \ref{figure5}a) is nearly pinned to this line (the other
canonical bcc structure with symmertry F\={4}3m lies much higher in energy). 
On the other hand, the $E^{\sub{(fm)}}_f$ curves originating from the pentahedron 
CEs run clearly below the B2-NiAl$-$B2-FeAl tie line and display a convex behavior
close to NiFeAl$_2$. Hence from the latter, an ordered ground state seems likely 
at this stoichiometry.  Two structures, here named ``G1'' and ``G2'' (see Fig. 
\ref{figure5}b,c), found by means of the higher pentahedron CEs, are located in the
favourable energy range. Actually, the G1 structure was identified as stable within
a first higher cluster approximation beyond the tetrahedron approximation. By 
refining the CE into the final P-CE, now with G1 belonging to the set of input 
structures of the respective SIM, we finally obtained the G2 structure as the 
stable ordered structure for this composition. Tab. \ref{table5} lists the
relevant data for the competing structures at NiFeAl$_2$. The structures found
from the CE investigation are by construction fixed on the rigid bcc parent 
lattice, since in the CE the total energy including the relaxation energy is 
represented as a function of configurations on the undistorted parent lattice. 
In reality, the structures G1 and G2 possess tetragonal symmetry, and only for the
G2 structure no forces appear in the bcc configuration. The G1 structure consists 
of alternating [100] Ni, Al and Fe planes. When actually relaxing G1 within the 
MBPP code, the Al planes are shifted towards the 
Fe planes. Still, structurally relaxing the G1 structure does not shift its 
formation energy below the value for G2 within the full electronic structure
method. The corresponding formation energies obtained via the P-CE and rP-CE are 
very close to the MBPP reference values and one yields the same energetic hierachy 
between the structures, showing the quantitative reliability of the constructed 
CEs. Concerning the curvature of the P-CE and rP-CE minimum formation energy 
curves, the overall progression between NiAl and FeAl excludes a tie line
construction that destabilizes the G2 structure. Moreover, it appears that at 
$c_{\sub{Fe}}$$=$0.125 there might be another pronounced ordering tendency,
indicated by a convex kink in the P-CE and rP-CE curves. This opens the 
possibility for an additional ordered state at Ni$_3$FeAl$_4$ within the given 
constraints. 
Concerning the behavior of the magnetic moment, the situation is delicate. As in 
DFT LDA/GGA ordered B2-NiAl is nonmagnetic and ordered B2-FeAl shows
ferromagnetism, an onset of the latter must appear on the NiAl$-$FeAl line. From 
the CE calculations, already close to B2-NiAl a fm moment should be expected. On 
the contrary, the MBPP calculations reveal no resulting magnetic moment for 
G1 and G2. Interestingly, close to the FeAl stoichiometry the magnetic moment 
strongly differs for different CEs. The complex interplay between structure and 
magnetism in the FeAl region is thus again manifest.
\subsection{Finite temperature phase diagram}
\begin{figure} % Fig 6: ternary phase diagram at 1250 K
\psfrag{Ni}{Ni}\psfrag{Fe}{Fe}\psfrag{Al}{Al}
\epsfig{file=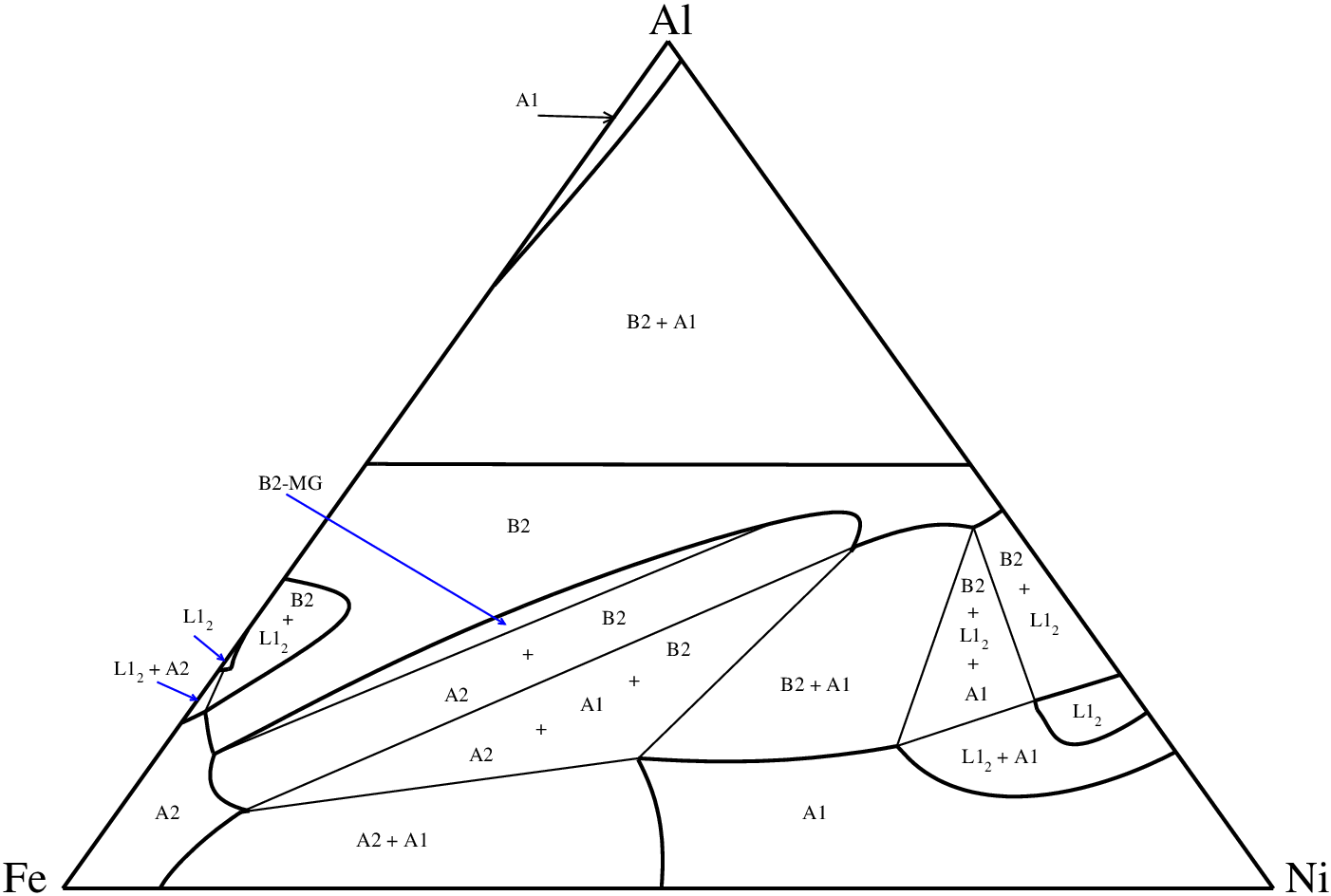,width=13cm}
\caption{Incoherent ab-initio phase diagram of Ni-Fe-Al on the bcc and fcc
lattice. Computed with the tetrahedron approximation to the CVM at $T$=1250 K.
\label{figure6}}
\end{figure}

\begin{figure} % Fig 7: ternary phase diagram at 500 K
\psfrag{Ni}{Ni}\psfrag{Fe}{Fe}\psfrag{Al}{Al}
\epsfig{file=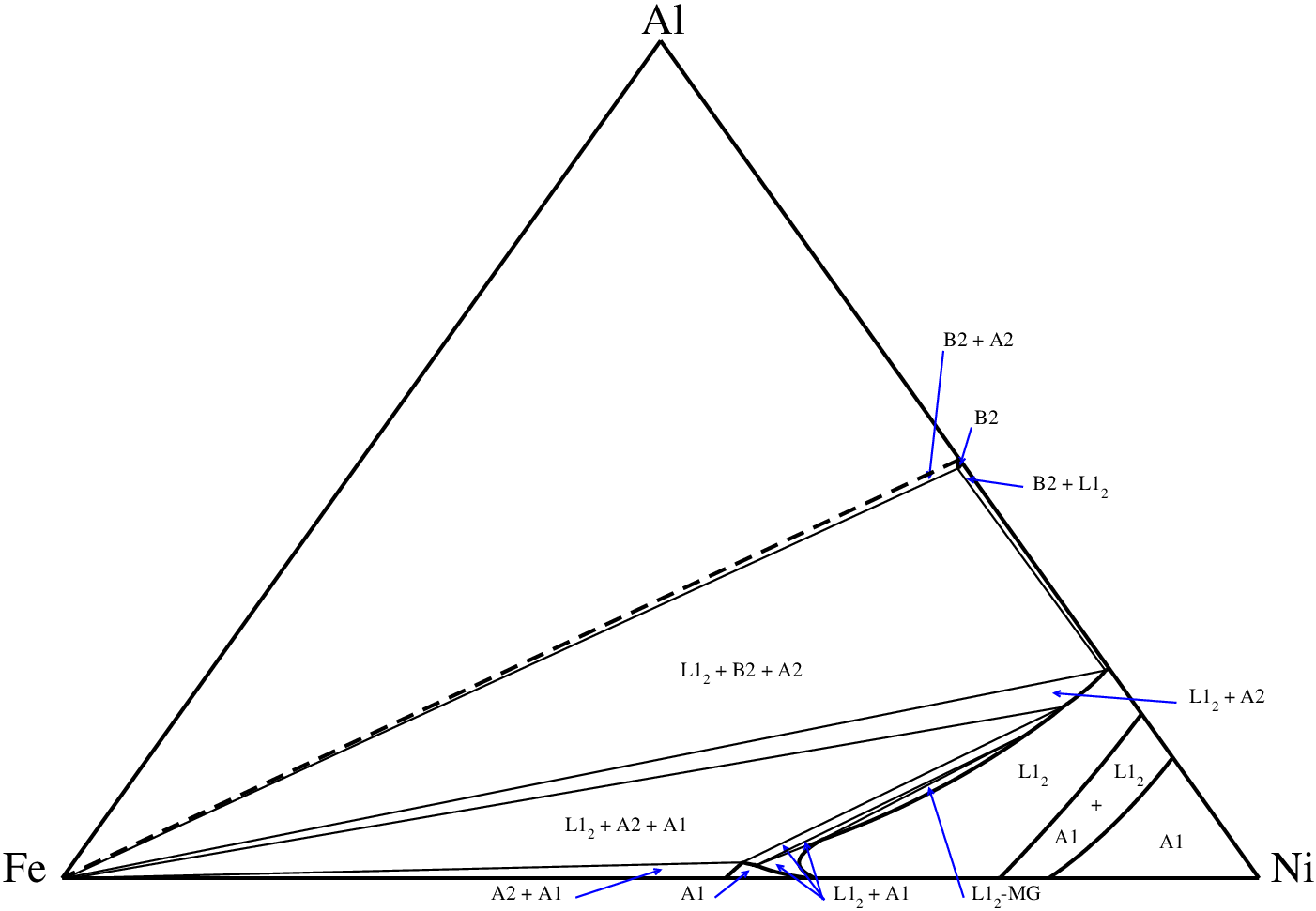,width=13cm}
\caption{Incoherent ab-initio phase diagram of Ni-Fe-Al on the bcc and fcc
lattice. Computed with the tetrahedron approximation to the CVM at $T$=500 K.
\label{figure7}}
\end{figure}

In this last section we want to discuss our result for the ternary Ni-Fe-Al phase 
diagram at finite temperature, investigated within the CVM using the ab-inito ECIs
for the energetics. The full ternary phase diagram is the result of the interplay 
of a large number of different mutually coupled degrees of freedom.
Because of the complexity of the problem we had to adopt approximations on several
levels. For completeness, we like to list these approximations in the following.
\begin{enumerate}
\item[a.] For the Fe-rich part of the phase diagram there is a very delicate
interplay between electronic correlations, magnetism and structure 
\cite{Cas73,Kat93,Ent93,Sch99,Lec04}, and for an accurate description of some of 
the binary Fe-Al subsystems it seems that one has to go beyond the LDA/GGA to the 
exchange-correlation energy of the DFT. In order to make the calculations for the 
whole ternary phase diagram feasible, we nevertheless adopted PBE-GGA for the 
construction of the ECIs.
\item[b.] We take into account the effect of magnetism by allowing for a possible
fm alignment of the magnetic moments, but the effect of thermal magnetic 
excitations is neglected.
\item[c.] We take into account the configurational entropy within the CVM, but
the contributions of phonons and single-electron excitations to the free energy
are neglected.
\item[d.] The effect of vacancies is neglected although vacancies are, at least, 
important for the Ni-Al phases on the bcc parent lattice \cite{Lec01}.
\item[e.] There are a few complex low-symmetry phases existing in the Al-rich
part of Ni-Fe-Al. We, however, allow only for phases on a bcc or fcc 
parent lattice.
\item[f.] For the internal energy and the entropy we employed the tetrahedron
approximation on the bcc and fcc parent lattice to the CVM. This allows us to 
consider only phases with up to four sublattices \cite{Col93}.
\item[g.] Due to the short range and the small number of degrees of freedom in the 
tetrahedron approximation, it seems not meaningful to include effects of local 
atomic relaxations, albeit they are surely important. Hence we did not include
such effects which would require to consider a correspondigly enlarged set of 
reference structures in a least-mean-squares fitted SIM.
\end{enumerate}
In view of these approximations we can not expect to be able to describe all the
details of the physics of the ternary phase diagram. However, we think that the
gross features and the qualitative trends are reproduced correctly. In this section
we only present our major results for the incoherent bcc/fcc phase diagram.
Additional considerations for $T$$\neq$0, also for the binary subsystems, can be 
found in \cite{Lec01,Diss03}.\\
We begin with a discussion of the ternary phase diagram computed at $T$=1250 K 
as shown in Fig. \ref{figure6}. Starting at low Al 
concentration, there is a wide heterogeneous B2$-$A1 phase mixture stabilized up 
to 50\% Al. Due to the low ordering energy in the upper half of the Gibbs triangle,
there are only a few complicated single-phase structures with rather big unit cells
(\cite{Bud92} and references therein). These structures can not be represented in 
the tetrahedron approximation, and thus are not considered in our investigation. 
A rather broad B2 phase is stabilized in the center of the Gibbs triangle. The 
other stable phases are also ternary extensions of the known stable binary phases.
Remarkably, the B2 phase varies continuously from NiAl to FeAl, i.e., B2-NiAl and
B2-FeAl are soluble into one another over a wide composition range. However, our
calculations also reveal a miscibility gap (MG) within the ternary B2-(Ni,Fe)Al 
phase along the Fe$-$NiAl direction which is already known from 
experiments \cite{Riv80,Liu92}. Our approach 
allows for the identification of the two different B2 phases involved in the MG. 
The first B2 phase is mainly oriented along NiAl$-$FeAl and can be defined by 
stating that the occupation of the Al sublattice of the B2 structure is strongly 
dominated by the Al atoms, whereas the substitution processes of Ni and Fe takes 
place solely via the transition metal sublattice. This type of microstructure was 
already found in our ground state calculations along NiAl$-$FeAl (section
\ref{tersys}). 
The second B2 phase develops for higher Fe concentration along A2-Fe$-$B2-NiAl. 
There, the majority Fe atoms now also manifestly replace Al atoms on their 
sublattice, i.e., the Fe atoms are not anymore mainly restricted to the transition
metal 
sublattice. The latter type of B2 phase obviously triggers the continuous B2$-$A2 
transition towards the Fe corner. The location of the MG is close to the 
region were the overall stability of the parent bcc lattice is rather delicate
(see Fig. \ref{figure3}b). Hence, the appearance of the MG may be related to the
exisiting of the two distinct minima, i.e., one at the NiAl stoichiometry and the 
other in the Fe corner, in the formation energy landscape for homogeneous
phases on the bcc parent lattice, as discussed in section \ref{tersys}. In 
connection to 
this it seems that by starting from B2-FeAl the sole transformation of the Fe
sublattice into an Ni sublattice when going towards the NiAl stoichiometry, as 
well as the sole vanishing of the Al sublattice when going towards the Fe corner,
can be accomplished continuously. In contrast, the system gets somehow 
``frustrated'' when the two B2 sublattices have to be transformed simultaneously,
as it happens when going along the FeAl$-$Ni line. Then it is thermodynamically 
more favourable to eventually open the MG and stabilize a heterogeneous mixture of
two B2 phases. Finally, by continuing along the FeAl-Ni line this MG also breaks 
down and a full B2$-$A2 mixture becomes stable. Note that a ``stoichiometric 
splitting'',
i.e., the formation of two lines of stability for B2-(Ni,Fe)Al is also described 
in the experimental work of Tan \textsl{et al.} \cite{Tan01}. We think that the
confirmation of the MG is a very big success of our ab-initio statistical mechanics
in view of the fact that it does not involve any fit parameter.\\
The topology of the obtained ab-initio phase diagram at $T$=1250 K matches nicely 
the one of the experimental phase diagram around 670 K published in the book of 
Sauthoff \cite{Sau95}. We want to remark on two specific problems concerning the 
latter comparison. First, due to the wrong stabilization of the L1$_2$ structure 
instead of the D0$_3$ structure for Fe$_3$Al at $T$=0K in PBE-GGA, the 
L1$_2$-Fe$_3$Al phase also appears as a stable phase in our calculated ternary 
phase diagram. Second, in nature fcc-Fe is stabilized against bcc-Fe at 
$T$=1250 K. The corresponding allotropic transformation, mainly driven by phonons,
can of course not be reproduced by our approach. The matching of our high-$T$ 
phase diagram with the experimental one at much lower temperatures may be
due to the limitations of the tetrahedron approximation and the fact that DFT 
calculations often tend to overestimate the formation energy. Our phase diagram at
1250 K does not include the L1$_2$-Ni$_3$Fe phase although we have identified the
corresponding configuration as a ground-state structure. This is due to the fact 
that the binary Ni$_3$Fe phase already disorders at 660 K in our CVM examinations 
of binary Ni-Fe \cite{Diss03}, and it is surely not expected that a ternary 
continuation of Ni$_3$Fe leads to a tremendeous increase of the order-disorder 
transition temperature $T^{\sub{(dis)}}_{\sub{Ni$_3$Fe}}$. Thus in order to 
investigate the phase stability
of possible ternary Ni$_3$Fe we calculated the incoherent Ni-Fe-Al phase diagram 
also at 
500 K (Fig. \ref{figure7}), but only in the Ni-Fe-NiAl triangle of the full Gibbs
triangle. In addition to the already exisiting phases at 1250 K, indeed there is
now also a Ni$_3$Fe phase with L1$_2$ symmetry. Surprisingly, a ternary 
Ni$_3$Fe phase was not indicated in the experimental phase diagram around 670 K 
from Ref. \cite{Sau95}, although the experimental value of 
$T^{\sub{(dis)}}_{\sub{Ni$_3$Fe}}$ mounts up to 790 K \cite{Swa91}. In our 
calculation Ni$_3$Fe and Ni$_3$Al are well soluble into one another over a wide
composition range, in analogy to the case of B2. Moreover, within the given 
approximations there is also a miscibility gap (MG) existing in this generic 
L1$_2$ phase. Further calculations revealed that this L1$_2$-MG has vanished at 
600 K, thus the MG should be confined to the low temperature regime. Experimental 
investigations to test these latter low temperature results are highly desirable.
\subsection{Conclusions}
In conclusion, we have presented a first-principles modelling of the
technologically important Ni-Fe-Al system. The emphasis of our work is on the 
completeness of the description rather then on specific details of this complex
ternary system. Hence, approximations, both on a physical and numerical level,
had to be applied to render the approach feasible. Although some of these 
approximations surely have to be removed in detailed analyses of certain
properties of Ni-Fe-Al, we think that our approach yields important
information for an identification of the relevant physical processes and 
interactions that have dominant influence in this intermetallic system. 
For instance, we have shown that the influence of magnetism is not negligible
for a vast composition range. The atomic ordering in the ternary B2 phase is
another interesting result, not only from the point of view of basic research 
but also concerning the manufacturing of new alloys with varying Fe content. 
Generally, in spite of the complexity of the problem, our rather simple approach 
yields encouraging agreement with the real Ni-Fe-Al system. Of 
course, more specialized theoretical and experimental work is 
needed. Thus, we hope to generally stimulate further reseach on this and other 
multicomponent intermetallic systems, since they provide a fascinating
variety of interacting physical processes due to the enlarged number of 
degrees of freedom. After all, such systems are generally of higher 
technological relevance than the more restricted binary systems.
\ack{This work was supported by the Deutsche Forschungsgemeinschaft (DFG) under 
the project No. FA 196/9-1.}

% \appendix

% \section{}
% \label{}

\newpage
\begin{table}[h]
\begin{tabular}{l|r|r|r|r}\hline
  &  &  &  &  \\ 
\bw{structure} & \bw{$a_{eq}$ $\left[\mbox{\AA}\right]$} & 
\bw{$B$ [Mbar]} & \bw{$E_f$ $\left[\frac{\mbox{meV}}{\mbox{atom}}\right]$} 
& \bw{$M$ $\left[\mu_{\mbox{\tiny{B}}}\right]$} \\ \hline\hline
                         & 2.812 & 1.891 &  119.6 & \\[-0.2cm] 
\bw{bcc-Ni}              & 2.821 & 1.825 &   93.5 & 0.56 \\[-0.15cm] \hline
                         & 2.782 & 2.666 &  596.2 & \\[-0.2cm]
\bw{bcc-Fe$^*$}          & 2.864 & 1.595 &    0.0 & 2.27 \\[-0.15cm] \hline
                         & 3.230 & 0.677 &   93.7 & \\[-0.2cm]
\bw{bcc-Al}              & 3.230 & 0.677 &   93.7 & 0.00\\[-0.15cm] \hline
                         & 2.797 & 2.214 &  445.7 & \\[-0.2cm] 
\bw{B2-NiFe}             & 2.877 & 1.673 &   58.6 & 3.58 \\[-0.15cm] \hline
                         & 2.898 & 1.546 & -676.8 & \\[-0.2cm] 
\bw{B2-NiAl$^*$}         & 2.898 & 1.546 & -676.8 & 0.00\\[-0.15cm] \hline
                         & 2.872 & 1.762 & -288.7 & \\[-0.2cm] 
\bw{B2-FeAl$^*$}         & 2.879 & 1.550 & -311.3 & 0.73 \\[-0.15cm] \hline
                         & 2.797 & 2.254 &  413.4 & \\[-0.2cm]
\bw{B32-Ni$_2$Fe$_2$}    & 2.859 & 1.755 &    7.2 & 6.70 \\[-0.15cm] \hline
                         & 2.929 & 1.433 & -357.3 & \\[-0.2cm]
\bw{B32-Ni$_2$Al$_2$}    & 2.929 & 1.433 & -357.3 & 0.00\\[-0.15cm] \hline
                         & 2.899 & 1.698 &   20.8 & \\[-0.2cm]
\bw{B32-Fe$_2$Al$_2$}    & 2.941 & 1.406 & -211.6 & 3.95 \\[-0.15cm] \hline
                         & 2.806 & 2.053 &  307.7 & \\[-0.2cm] 
\bw{D0$_3$-Ni$_3$Fe}     & 2.842 & 1.813 &   11.9 & 4.56 \\[-0.15cm] \hline
                         & 2.789 & 2.445 &  521.9 & \\[-0.2cm] 
\bw{D0$_3$-NiFe$_3$}     & 2.882 & 1.738 &   10.1 & 8.85 \\[-0.15cm] \hline
                         & 2.847 & 1.740 & -389.9 & \\[-0.2cm]
\bw{D0$_3$-Ni$_3$Al}     & 2.847 & 1.740 & -389.9 & 0.00 \\[-0.15cm] \hline
                         & 3.059 & 1.013 &  -97.7 & \\[-0.2cm] 
\bw{D0$_3$-NiAl$_3$}     & 3.059 & 1.013 &  -97.7 & 0.00 \\[-0.15cm] \hline
                         & 2.821 & 2.181 &  106.1 & \\[-0.2cm] 
\bw{D0$_3$-Fe$_3$Al$^*$} & 2.892 & 1.510 & -201.0 & 6.35 \\[-0.15cm] \hline
                         & 2.990 & 1.274 &  -13.1 & \\[-0.2cm] 
\bw{D0$_3$-FeAl$_3$}     & 2.989 & 1.196 &  -13.1 & 0.00 \\[-0.15cm] \hline
                         & 2.850 & 1.836 &  -98.7 & \\[-0.2cm] 
\bw{L2$_1$-Ni$_2$FeAl}   & 2.886 & 1.631 & -345.4 & 3.39 \\[-0.15cm] \hline
                         & 2.834 & 1.991 &   46.9 & \\[-0.2cm] 
\bw{L2$_1$-NiFe$_2$Al}   & 2.890 & 1.437 &  -98.6 & 4.88 \\[-0.15cm] \hline
                         & 2.878 & 1.662 & -492.2 & \\[-0.2cm] 
\bw{L2$_1$-NiFeAl$_2$}   & 2.881 & 1.492 & -494.3 & 0.36 \\[-0.15cm] \hline
                         & 2.844 & 1.847 & -130.0 & \\[-0.2cm] 
\bw{F\={4}3m-Ni$_2$FeAl} & 2.869 & 1.635 & -279.2 & 2.88 \\[-0.15cm] \hline
                         & 2.834 & 2.011 &    1.7 & \\[-0.2cm]  
\bw{F\={4}3m-NiFe$_2$Al} & 2.883 & 1.677 & -299.8 & 4.90 \\[-0.15cm] \hline
                         & 2.918 & 1.545 & -120.9 & \\[-0.2cm] 
\bw{F\={4}3m-NiFeAl$_2$} & 2.956 & 1.276 & -178.7 & 2.26 \\[-0.15cm] \hline\hline
\end{tabular}
\caption{Structural data for the bcc structures used in the SIM for the 
tetrahedron approximation (see text).
\label{table1}}
\end{table}
%%%%%%%%%%%%%%%%%%%%%%%%%%%%%%
\begin{table}
\begin{tabular}{l|r|r|r|r}\hline
  &  &  &  &  \\ 
\bw{structure} & \bw{$a_{eq}$ $\left[\mbox{\AA}\right]$} & 
\bw{$B$ [Mbar]} & \bw{$E_f$ $\left[\frac{\mbox{meV}}{\mbox{atom}}\right]$} 
& \bw{$M$ $\left[\mu_{\mbox{\tiny{B}}}\right]$} \\ \hline\hline
                         & 3.536 & 1.940 &   65.9 & \\[-0.2cm] 
\bw{fcc-Ni$^*$}          & 3.544 & 1.892 &    0.0 & 0.64 \\[-0.15cm] \hline
                         & 3.478 & 2.790 &  286.2 & \\[-0.2cm] 
\bw{fcc-Fe}              & 3.668 & 1.643 &  125.6 & 2.65 \\[-0.15cm] \hline
                         & 4.036 & 0.755 &    0.0 & \\[-0.2cm] 
\bw{fcc-Al$^*$}          & 4.036 & 0.755 &    0.0 & 0.00 \\[-0.15cm] \hline
                         & 3.509 & 2.320 &  287.9 & \\[-0.2cm] 
\bw{L1$_0$-NiFe}         & 3.594 & 1.788 &  -77.8 & 6.69 \\[-0.15cm] \hline
                         & 3.671 & 1.492 & -531.1 & \\[-0.2cm] 
\bw{L1$_0$-NiAl}         & 3.671 & 1.492 & -531.1 & 0.00 \\[-0.15cm] \hline
                         & 3.659 & 1.700 &  -69.6 & \\[-0.2cm]
\bw{L1$_0$-FeAl}         & 3.704 & 1.391 & -270.3 & 3.54 \\[-0.15cm] \hline
                         & 3.524 & 2.117 &  214.5 & \\[-0.2cm] 
\bw{L1$_2$-Ni$_3$Fe$^*$} & 3.571 & 1.864 &  -98.3 & 4.91 \\[-0.15cm] \hline
                         & 3.493 & 2.543 &  310.4 & \\[-0.2cm] 
\bw{L1$_2$-NiFe$_3$}     & 3.630 & 1.523 &   29.8 & 8.56 \\[-0.15cm] \hline
                         & 3.581 & 1.763 & -434.8 & \\[-0.2cm] 
\bw{L1$_2$-Ni$_3$Al$^*$} & 3.582 & 1.757 & -441.0 & 0.74 \\[-0.15cm] \hline
                         & 3.843 & 1.100 & -230.0 & \\[-0.2cm]
\bw{L1$_2$-NiAl$_3$}     & 3.843 & 1.100 & -230.0 & 0.00 \\[-0.15cm] \hline
                         & 3.567 & 2.137 &  174.3 & \\[-0.2cm] 
\bw{L1$_2$-Fe$_3$Al}     & 3.669 & 1.680 & -222.0 & 6.99 \\[-0.15cm] \hline
                         & 3.793 & 1.275 & -105.3 & \\[-0.2cm] 
\bw{L1$_2$-FeAl$_3$}     & 3.797 & 0.988 & -105.3 & 0.26 \\[-0.15cm] \hline
                         & 3.581 & 1.865 & -158.5 & \\[-0.2cm] 
\bw{P4/mmm-Ni$_2$FeAl}   & 3.617 & 1.664 & -325.1 & 3.00 \\[-0.15cm] \hline
                         & 3.573 & 2.004 &   17.5 & \\[-0.2cm] 
\bw{P4/mmm-NiFe$_2$Al}   & 3.646 & 1.626 & -263.7 & 5.08 \\[-0.15cm] \hline
                         & 3.664 & 1.598 & -309.6 & \\[-0.2cm] 
\bw{P4/mmm-NiFeAl$_2$}   & 3.691 & 1.414 & -392.5 & 1.78 \\[-0.15cm] \hline\hline
\end{tabular}
\caption{Structural data for the fcc structures used in the SIM for the 
tetrahedron approximation (see text).\label{table2}}
\end{table}
%%%%%%%%%%%%%%%%%%%%%%%%%%%%%%
\begin{table}
\begin{center}
\begin{tabular}{l|r r r r|r r r r}\hline
             & & bcc&       &  &  & fcc &       & \\ \hline
cluster $\alpha$ & $\Theta_2$ & & $\Theta_1$ & & $\Theta_2$ & & $\Theta_1$ &\\ 
$m$       & $K^{\sub{(nm)}}_{\alpha m}$ & 
$K^{\sub{(fm)}}_{\alpha m}$ & 
$K^{\sub{(nm)}}_{\alpha m}$ & $K^{\sub{(fm)}}_{\alpha m}$ & 
$K^{\sub{(nm)}}_{\alpha m}$ & $K^{\sub{(fm)}}_{\alpha m}$ & 
$K^{\sub{(nm)}}_{\alpha m}$ & $K^{\sub{(fm)}}_{\alpha m}$    \\ 
\hline\hline
empty      &  -29.0 & -203.1 &   93.7 &   93.7 &  -89.1 & -261.6 &    0.0 &  
  0.0 \\
point     &    &  &  &  &  & &  &   \\
1         & -243.8 &  -85.7 & -169.2 & -169.2 & -258.7 & -128.8 & -249.4 & 
-249.4  \\
2         & -330.3 &  -81.4 & -596.4 & -596.4 & -308.4 & -118.4 & -670.6 & 
-670.6  \\
NN pair   &  &  &  &  &  & &  &   \\
11        &  -13.6 &   -9.8 &  -23.7 &  -52.9 &   -8.2 &    2.2 &    4.6 &   
-4.1  \\
12        &   -9.5 &  -35.9 & -104.5 &   11.8 &  -26.1 &  -13.9 & -106.1 &   
-5.7  \\
22        &   92.1 &   71.8 &   59.9 &  -27.1 &   48.5 &   47.3 &   30.3 &  
-61.3  \\
NNN pair  &  &  &  &  &  &  &  &  \\
11        &   -2.3 &    5.3 &    3.2 &    0.1 &     -  &      - &    -   &    
-    \\
12        &   -4.7 &    8.4 &  -73.0 &  -65.4 &     -  &      - &    -   &    
-    \\
22        &   10.6 &    8.9 & -188.7 & -193.1 &     -  &      - &    -   &    
-    \\
triangle   &  &  &  &  &  & &  &   \\
111       &   -1.0 &    0.8 &   -3.6 &    0.7 &   -1.2 &    1.1 &   -5.1 &   
-2.2  \\
112       &   -2.1 &   -4.8 &   -3.6 &   23.0 &    2.8 &   -7.6 &  -29.3 &   
-7.8  \\
121       &    0.0 &   -0.3 &   -4.9 &    5.1 &   -    &   -    &    -   &    
-    \\
122       &    3.3 &    2.2 &   46.0 &   20.7 &   16.4 &   -0.4 &  105.4 &   
49.5  \\
212       &    1.7 &    1.8 &   23.3 &   14.3 &   -    &   -    &    -   &    
-    \\
222       &   -3.1 &   -3.6 &  147.6 &  141.1 &    2.3 &   -1.5 &  163.3 &  
191.6  \\
tetrahedron &  &  &  &  &  &  &  &  \\
1111      &   -0.3 &    0.7 &   -0.6 &    1.6 &   -0.3 &   -0.9 &   -0.6 &   
-2.1  \\
1112      &   -1.3 &   -0.6 &    5.2 &    2.4 &   -4.3 &   -6.4 &   16.7 &   
25.0  \\
1122      &    2.3 &   -1.7 &   15.6 &  -11.8 &    9.1 &   14.2 &   61.6 &   
96.1  \\
1212      &    1.1 &   -1.0 &    7.2 &   -6.4 &   -    &   -    &    -   &    
-    \\
1222      &    3.6 &    1.1 &  -42.1 &  -12.9 &    2.6 &    8.9 &  -30.2 & 
-103.9  \\
2222      &   -4.4 &   -4.0 &  -88.4 &  -80.4 &  -13.7 &  -13.2 & -278.2 & 
-266.4  \\ 
\hline\hline
\end{tabular}
\caption{Ternary CE of the formation energy in tetrahedron approximation. ECIs
$K_{\alpha m}$ in meV/atom.\label{table3}}
\end{center}
\end{table}
%%%%%%%%%%%%%%%%%%%%%%%%%%%%%%
\begin{table}[b]
\begin{tabular}{l|r r|r r}\hline
               & bcc&    & fcc &  \\ 
\bw{EPI}       & nm & fm & nm  &fm   \\ \hline\hline
$W^{(1)}_{NiFe}$ & -20.4 & -14.7 & -12.3 &  3.3 \\  
$W^{(2)}_{NiFe}$ &  -3.5 &   8.0 &  -    &  -  \\ \hline 
$W^{(1)}_{NiAl}$ & 110.9 & 123.7 &  85.4 & 72.1 \\ 
$W^{(2)}_{NiAl}$ &  17.2 &   1.1 &  -    &  -  \\ \hline  
$W^{(1)}_{FeAl}$ &  86.2 &  30.5 &  17.6 & 36.0 \\ 
$W^{(2)}_{FeAl}$ &   5.0 &  22.9 &  -    &  -  \\ \hline\hline
\end{tabular}
\caption{Effective quasibinary pair interactions in the Ni-Fe-Al system
(in meV/atom).\label{table4}}
\end{table}
%%%%%%%%%%%%%%%%%%%%%%%%%%%%%%
\begin{table}[h]
\begin{center}
\begin{tabular}{l|r|r|l l l|r r r}\hline
 &  &  &  &  &  &  &  &   \\
\bw{structure} & \bw{$a$ [a.u.]} & \bw{$B$ [Mbar]} &
\bw{$E_{form}$} & \bw{$\left[\mev\right]$} & & \bw{M [$\mub$]} &  & \\
 &  &  & MBPP & P-CE & rP-CE & MBPP & P-CE & rP-CE \\ \hline\hline
L2$_1$    & 5.445 & 1.492 & -494.3 & -494.3 & -494.3 &  0.36 & 0.36 & 0.36 \\
G1        & 5.443 & 1.586 & -519.4 & -503.2 & -517.4 &  0.00 & 0.36 & 0.16 \\
G2        & 5.433 & 1.530 & -528.8 & -523.9 & -524.2 &  0.00 & 0.48 & 0.48 \\ 
\hline
NiAl$-$FeAl &       &       & -494.1 &        &        &       &      &      \\ 
\hline\hline
\end{tabular}
\caption{Structural data for the competing structures at NiFeAl$_2$. The MBPP 
value $E^{\sub{(fm)}}_f$ for the G2 structure belongs to the structurally relaxed 
geometry.\label{table5}}
\end{center}
\end{table}

%%%%%%%%%%%%%%%%%%%%%%%%%%%%%%
%%%%%%%%%%%%%%%%%%%%%%%%%%%%%%

\newpage
%%%%%%%%%%%%%%%%%%%%%%%%%%%%%%
\newpage
%%%%%%%%%%%%%%%%%%%%%%%%%%%%%%
\newpage
%%%%%%%%%%%%%%%%%%%%%%%%%%%%%%
\newpage
%%%%%%%%%%%%%%%%%%%%%%%%%%%%%%
\newpage
%%%%%%%%%%%%%%%%%%%%%%%%%%%%%%
\newpage
%%%%%%%%%%%%%%%%%%%%%%%%%%%%%%
\newpage
%%%%%%%%%%%%%%%%%%%%%%%%%%%%%%
\end{document}